\documentclass[
superscriptaddress,
 amsmath,amssymb,
 aps,
 pra,
  twocolumn,
floatfix,
]{revtex4-2}

\usepackage{graphicx}
\usepackage{dcolumn}
\usepackage{bm}
\usepackage{xcolor}
\newcommand{\dd}{\mathrm{d}}

\begin{document}

\preprint{APS/123-QED}

\title{Optical Implementation of Equilibrium Propagation \\ Using Spatial
Photonic Ising Machines}

\author{Dimitri Vanden Abeele}
\thanks{These authors contributed equally to this work. Corresponding author: dimitri.vanden.abeele@ulb.be}

\affiliation{Laboratoire d'Information Quantique C.P. 224, Universit\'e Libre de Bruxelles (ULB), Brussels, Belgium}

\author{Daniele Veraldi}
\thanks{These authors contributed equally to this work. Corresponding author: dimitri.vanden.abeele@ulb.be}
\affiliation{Dipartimento di Fisica, Sapienza Universit\`a di Roma, Rome, Italy}

\author{Davide Pierangeli}
\affiliation{Dipartimento di Fisica, Sapienza Universit\`a di Roma, Rome, Italy}

\author{Claudio Conti}
\affiliation{Dipartimento di Fisica, Sapienza Universit\`a di Roma, Rome, Italy}

\author{Serge Massar}
\affiliation{Laboratoire d'Information Quantique C.P. 224, Universit\'e Libre de Bruxelles (ULB), Brussels, Belgium}

\date{\today}

\begin{abstract}
Equilibrium Propagation offers a compelling alternative to traditional machine learning for training energy-based networks. Here we demonstrate a hybrid optical-digital implementation of EP using a Spatial Photonic Ising Machine (SPIM). The SPIM exploits the gauge transformation method to optically encode both continuous neuron states and rank‑1 binary trainable patterns as phase modulations via a spatial light modulator, with inference realized using a finite difference scheme. The experimental system is evaluated on the Wine classification dataset. The potential of this approach, including the use of continuous couplings and structured coupling matrices, is evaluated numerically on the more complex MNIST dataset. Our work provides a concrete pathway toward energy-efficient physical implementations of Equilibrium Propagation.

\end{abstract}

\maketitle


\section{Introduction}
Hardware implementations of machine learning algorithms are receiving increasing attention because of their potential for low energy consumption, particularly in photonics \cite{zhang2021efficient, wright2022deep, pai2023experimentally, xue2024fully, spall2025training}. However, state-of-the-art digital networks are trained using error backpropagation, an algorithm difficult to map onto analog physical systems. Although \textit{in situ} optical backpropagation has recently been demonstrated \cite{pai2023experimentally, spall2025training}, it relies on bidirectional light propagation and precise optical phase measurements, which impose strict hardware constraints.

These implementation challenges have renewed interest in contrastive energy-based approaches
\cite{hopfield1982neural, ackley1985learning, hinton2025nobel}. Rather than adapting physical systems to implement backpropagation, these approaches seek to exploit the dynamics of the underlying substrate.
Among these methods, the recently introduced Equilibrium Propagation (EP) \cite{scellier2017equilibrium} provides an exact supervised learning framework that trains energy-based networks by exploiting their intrinsic relaxation toward equilibrium, without the need for explicit error backpropagation. In EP, gradients are estimated by comparing the system's equilibrium states when output neurons are nudged toward a desired target versus when they are free. Thus, a physical implementation requires two basic ingredients: a system governed by a tunable energy functional, and a mechanism to perturb the outputs of the system corresponding to the prediction (see Ref.\ \cite{momeni2025training} for a recent review on physical learning).

Several extensions of EP have been proposed to eliminate nudged-state storage \cite{ernoult2020equilibrium, falk2025temporal}, derive agnostic updates for black-box energies \cite{scellier2022agnostic}, replace nudging with coupling via Coupled Learning \cite{stern2021supervised}, and generalize to spiking neurons \cite{martin2021eqspike, o2019training}, non-conservative dynamics \cite{scellier2018generalization, scurria2026equilibrium, stern2026contrastive},  stochastic systems \cite{scellier2017equilibrium, massar2025equilibrium} and dynamical systems with time-varying inputs \cite{massar2025EPlagrangian, pourcel2025lagrangian, berneman2026equilibrium,  stern2026contrastive}. Experimental realizations, however, remain scarce. Recent examples include the EP-inspired MADEM algorithm on memristor crossbars \cite{yi2023activity}, Coupled Learning in self-adjusting electrical circuits \cite{dillavou2022demonstration, dillavou2024machine} and elastic networks \cite{altman2024experimental}, and classical Ising models trained on a D-Wave quantum annealer \cite{laydevant2024training}. In these setups, the physical hardware either handles inference alongside an external (digital) update loop \cite{yi2023activity, altman2024experimental, laydevant2024training}, or operates as a fully autonomous, processor-free learning system \cite{dillavou2022demonstration, dillavou2024machine}. These implementations are further complemented by \textit{in silico} simulations, such as on analog circuits \cite{kendall2020training, oh2023memristor} and coupled oscillators \cite{wang2024training, rageau2025training}.

To address the need for parallel, energy-efficient physical systems, Spatial Photonic Ising Machines (SPIMs) offer a promising platform.
SPIMs were originally designed to solve quadratic unconstrained binary optimization (QUBO) problems by minimizing the energy of an equivalent Ising model \cite{wang2025efficient, NPmappedtoIsing, kalinin2020complexity}. 
SPIMs exploit the parallelism of Spatial Light Modulators (SLMs) to encode spins and couplings into the spatial degrees of freedom of a coherent optical field \cite{pierangeli2019large}, computing the total Ising energy via free-space propagation and a single intensity measurement. They are potentially offering advantages in energy, scalability, and/or speed (see Ref.~\cite{brunner2025roadmap} for other approaches to photonic computing). 

Here we demonstrate that a SPIM can be used to realize a hybrid optical-digital implementation of EP. We use the fully programmable SPIM previously reported in \cite{veraldi2024fully} to optically evaluate most of the terms of the Hamiltonian, as well as its gradients. The remaining terms are evaluated numerically and then summed. We successfully demonstrate the system on the Wine classification dataset \cite{wine_109}.
This approach shares similarities with Ref.~\cite{yamashita2023low}, which used a hybrid SPIM-digital method to train a Boltzmann Machine on binary MNIST images.

In their original formulation, SPIMs encode binary variables (spins). This limitation can be circumvented to some extent using the gauge transformation introduced in \cite{fang2021experimental}. Here we use this gauge transformation to represent continuous neurons. However, most of the  trainable parameters are then restricted to binary values. This does not prevent training on the  Wine dataset, but limits performance on more complex datasets. In our numerical study, we show that a future SPIM with continuous parameters can give good results on MNIST \cite{lecun1998gradient}. We also numerically investigate deeper architectures and the impact of quantization errors.

The present work thus broadens the potential applications of SPIMs and provides a basis for developing analog, low-energy neuromorphic computing systems.

\section{Theory}
\begin{figure}
    \centering
    \includegraphics[width=1\linewidth]{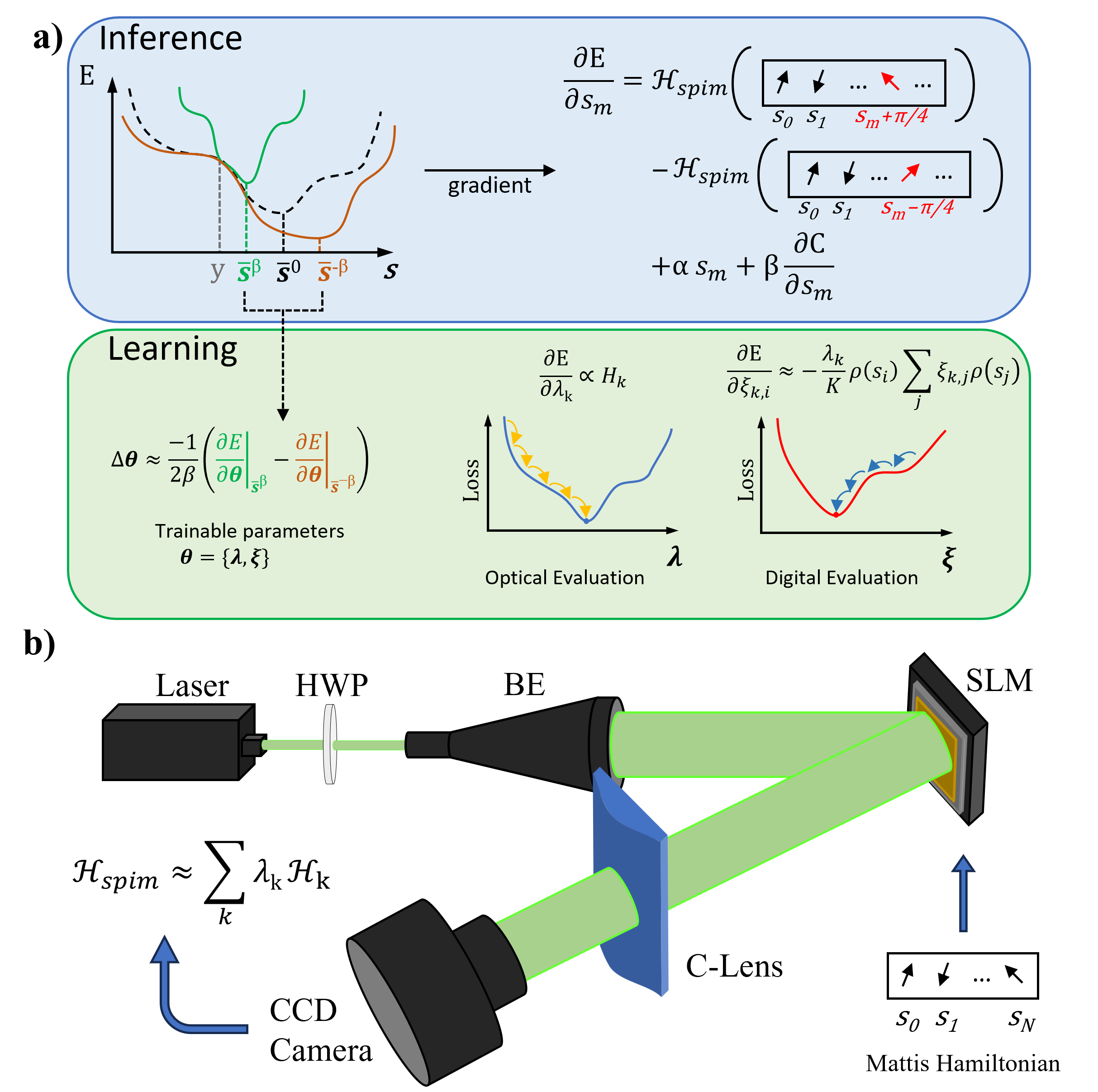}
    \caption{(a) Equilibrium Propagation implemented using SPIM. EP follows a bi-level optimization. 
    During inference the energy is minimized using a combination of optical measurements and digital terms. 
    During learning the  energy landscape is modified to include the cost, corresponding to the additional force 
     $\pm \beta \partial C / \partial s_m$. The trainable parameters $\theta=\{\lambda, \xi\}$ are updated according to how the variables conjugate to $\theta$ respond, at equilibrium, to the nudging. The variables $\partial E / \partial \lambda_k$ are evaluated optically, while the variables 
     $\partial E / \partial \xi_{k,i}$ are evaluated digitally.
    (b) Experimental setup. The SPIM evaluates the value of Mattis Hamiltonians  $\mathcal{H}_k \propto  \frac{1}{2}  \sum_{i,j=1}^{N}  \xi_{k,i} \sin(s_i) \xi_{k,j} \sin(s_j) $.   
    HWP: Half-wave plate, BE: Beam expander, C-Lens: Cylindrical lens, CCD: Charge-coupled device.}
    \label{fig:SpimSetup}
\end{figure}

Equilibrium Propagation (EP), summarized in Fig.~\ref{fig:SpimSetup}.a, trains dynamical systems that relax to the ground state $\bar{\bm{s}}$ of an energy functional $E$, given clamped inputs $\bm{u} \in \mathbb{R}^{N_\text{i}}$. The dynamical variables $\bm{s} \in \mathbb{R}^{N_\text{d}}$ decompose into hidden units $\bm{s}^\text{hid}$ and output units $\bm{s}^\text{out}$ weakly coupled to a target $\bm{y}$. The total energy reads:
\begin{equation}\label{eq:EnergyFunction}
    E = B(\bm{s}, \bm{u}) + I(\bm{s}) + \frac{\alpha}{2}\lVert\bm{s}\rVert^2 + \frac{\beta}{2} \lVert\bm{s}^\text{out} - \bm{y}\rVert^2 . 
\end{equation}
Here
$B(\bm{s}, \bm{u}) = - \rho(\bm{u})^\top  J^\text{in}   \rho(\bm{s})$ is the input bias, and
\begin{equation}
    \label{eq:quadraticInteraction}
    I(\bm{s}) = - \frac{1}{2} \rho(\bm{s})^\top {J}^\text{dyn} \rho(\bm{s}), 
\end{equation}
is the quadratic interaction, with ${J}^\text{dyn}$ a symmetric matrix. The element-wise nonlinearity $\rho(\bm{s})$, realized via optical interference, is taken to be a sine for $|s_i| \le \pi/2$ and to saturate to $\pm 1$ outside this interval, preventing the emergence of additional periodic local minima. The quadratic term $\frac{\alpha}{2}\lVert\bm{s}\rVert^2$, scaled by hyperparameter $\alpha$, bounds the state values to keep most units $s_i$ in the non-saturated range of the nonlinearity.
The learnable parameters are $\bm{\theta} = \{J^\text{in}, {J}^\text{dyn}\}$. 

EP learning follows a bi-level optimization: an inner relaxation minimizing $E$ (inference) during which   $\beta$ is set to zero and $\partial_{\bm{s}} E$ is integrated to a stationary state $\bar{\bm{s}}^0$, and an outer loop updating parameters $\bm{\theta}$  based on the equilibrium's sensitivity to $\beta$ (learning) \cite{zucchet2022beyond}. During the learning phase, two small nudges  ($\pm \beta$) are applied  to obtain  new equilibrium states $\bar{\bm{s}}^{\pm \beta}$ \cite{laborieux2021scaling}. The parameters are then updated according to
\begin{equation}\label{eq:EPUpdateRule}
    \Delta \bm{\theta} \propto -\frac{\dd}{\dd\beta} \frac{\partial E}{\partial \bm{\theta}} \approx \frac{1}{2\beta}\left( \left.\frac{\partial E}{\partial \bm{\theta}} \right|_{-\beta} - \left.\frac{\partial E}{\partial \bm{\theta}} \right|_{+\beta} \right).
\end{equation}
This update rule is spatially local, as the coupling decomposes into pairwise interactions.

Numerically, evaluating $\partial_{\bm{s}} I$ is the computational bottleneck, which requires $\mathcal{O}(N_\text{d}^2)$ matrix-vector multiplications at each step. 

To bypass this bottleneck, we use a SPIM. A SPIM provides a potentially fast method to optically evaluate  the Hamiltonian:
\begin{equation}
\mathcal{H}_\text{SPIM} \propto - \sum_{ij} {J}_{ij} \sin(s_i) \sin(s_j)
\label{Eq:SPIM}
\end{equation}
by taking linear combinations of rank-one Mattis problems \cite{yamashita2023low},
\begin{align}\label{Eq:Mattis}
   {J}_{ij} = \frac{1}{K}
    \sum_{k=1}^K \lambda_k \xi_{k,i} \xi_{k,j}.
\end{align}

The gradient of $\mathcal{H}_\text{SPIM}$ with respect to variable $s_m$ can be efficiently evaluated optically by a finite difference scheme:
\begin{equation}
    \begin{split}
      \left.  \frac{\partial \mathcal{H}}{\partial s_m}\right|_{\text{SPIM}} & = \left[\mathcal{H}_\text{SPIM}\left(s_{i\ne m}, s_m + \tfrac{\pi}{4}; {J}_{ij}\right) \right. \\ &\quad \left. - \mathcal{H}_\text{SPIM}\left( s_{i\ne m}, s_m - \tfrac{\pi}{4}; {J}_{ij}\right)\right]\ .
    \end{split} 
    \label{Eq:HSPIMpi/4}
\end{equation}

The trainable parameters $\bm{\theta}$ are the parameters $\{\lambda_k, \xi_{k,i}\}$ of the Mattis problems Eq. (\ref{Eq:Mattis}).
The variables conjugate to these parameters need to be evaluated in the nudged state, see Eq. (\ref{eq:EPUpdateRule}). 
The gradient with respect to $\lambda_k$ is simply the sum of the original Mattis problems:
\begin{equation}
    \left. \frac{\partial \mathcal{H}}{\partial \lambda_k}  \right|_{\text{SPIM}} \approx -\frac{1}{2K} \sum_{i,j}^N \xi_{k,i} \xi_{k,j} \rho(s_i) \rho(s_j).\label{eq:learnrule_lambda}
\end{equation}
which can be efficiently evaluated optically in a single shot.
The gradient with respect to the Mattis vectors $\xi_{k,i}$ 
\begin{equation}
     \frac{\partial I}{\partial \xi_{k,i}} \approx -\frac{\lambda_k}{K} \rho(s_i) \sum_j \xi_{k,j} \rho(s_j). \label{eq:learnrule_xi}
\end{equation}
could in principle also be evaluated optically (see Appendix~\ref{App-optical-grad}), but we evaluate it numerically to reduce the experiment run time. Evaluating it digitally is still efficient as it only needs to be computed once per pattern $k$ since the sum over $j$ is independent of $i$.

To understand how the SPIM is used, note that the system Eq. (\ref{eq:EnergyFunction}) can be rewritten by absorbing the input drive $B$ into the interaction term $I$. Defining the augmented state $\bm{x} = (\bm{u}, \bm{s})$ and the symmetric coupling
\begin{equation}
    J = \begin{bmatrix}
    0 & J^\text{in} \\
    (J^\text{in})^\top & J^\text{dyn} 
    \end{bmatrix},
    \label{Eq:SymmCoupling}
\end{equation}
one can rewrite $B(\bm{s}, \bm{u}) + I(\bm{s}) = 
- \frac{1}{2} \rho(\bm{x})^\top {J} \rho(\bm{x})$ 
which has the form of the SPIM Hamiltonian Eq. (\ref{Eq:SPIM}), and can therefore be evaluated optically.
In order to carry out the gradient descent step $s_m \leftarrow s_m - \epsilon \partial_{s_m} E$ (with $\epsilon$ a small learning rate), we combine  optically evaluated terms with simple digital operations:
\begin{align}
  \frac{\partial E}{\partial s_m} =& - \left( \left.(J^\text{in})^\top \rho(\bm{u}) \right)_m \rho'(s_m) \right|_{\text{SPIM}}  + \left.\frac{\partial I}{\partial s_m}\right|_{\text{SPIM}} \nonumber\\
        \ & + \alpha s_m + \beta (s_m - y_m) \delta_{m \in \text{out}}\ .
    \label{Eq:dEds}  
\end{align}
Finally, the gradients needed to update the parameters are evaluated optically and numerically, as described in Eqs. (\ref{eq:learnrule_lambda}, \ref{eq:learnrule_xi}), see Fig.~\ref{fig:SpimSetup}.a. The number $K$ of Mattis Hamiltonians is an important parameter determining the speed of the optical implementation. 
In Appendix~\ref{app:scaling}, we show that matrix $J$ in Eq.~\eqref{Eq:SymmCoupling} can be replaced with an equivalent coupling matrix with rank  upper bounded by the number $N_d$ of dynamical variables, implying a favorable scaling.  We also discuss in this same Appendix a hybrid scheme, similar to the one used in Ref. \cite{laydevant2024training}, in which the first term in Eq.~\eqref{Eq:dEds} is evaluated numerically, which could be useful for problems with large input dimension $N_i$.

After evaluation, the gradients are passed to standard optimizers such as stochastic gradient descent or ADAM \cite{kingma2014adam, ruder2016overview}. In the experimental implementation, because the parameters $\xi_{k,i} \in \{-1,1\}$  are binary, we use the Binary Optimizer (BOP) \cite{helwegen2019latent, laydevant2021training} which flips states based on an exponential moving average of the gradient governed by a threshold $\tau$ and adaptivity rate $\gamma$. 

The method Eq. (\ref{Eq:HSPIMpi/4}) to compute the gradient of the interaction terms in $E$  is convenient because it requires only two measurements of $\mathcal{H}_\text{SPIM}$. However, this method in fact computes the gradient of the energy Eq.~(\ref{eq:EnergyFunction}) for a slightly different coupling matrix $\tilde J_{ij}$ which differs from $J_{ij}$ by a rescaling of the diagonal elements. 
Therefore, Eq.~(\ref{Eq:HSPIMpi/4}) and Eqs.~(\ref{eq:learnrule_lambda},\ref{eq:learnrule_xi}) 
are not perfectly compatible. This is described in Appendix \ref{App-A} where numerical studies are  reported showing that this does not affect the performance of the system. For binary $\xi_{k,i}$ (which is the case  in the experiment) this rescaling  is essentially equal to a redefinition of hyperparameter $\alpha$.

\section{Experimental setup}
We experimentally demonstrate the implementation of EP
on the fully programmable SPIM described in Ref.~\cite{veraldi2024fully}, see Fig.~\ref{fig:SpimSetup}.b. Most SPIMs are limited to encoding rank-1 interaction matrices (\textit{i.e.}, Mattis models) 
\cite{pierangeli2019large, fang2021experimental, yamashita2023low}.
Taking linear combinations of  Mattis-type problems using Eq.~\eqref{Eq:Mattis} allows full programmability of the interaction matrix.
Inspired by the gauge transformation method of \cite{fang2021experimental}, we take the vectors in the Mattis problems to have binary coefficients $\xi_{k,i} \in \{-1,1\}$, and the dynamical variables to be continuous and belong to the range $s_i \in [-\pi/2, +\pi/2]$.

In practice, we use the Focal Plane Division (FPD) technique \cite{veraldi2024fully} to simultaneously compute the energies of multiple $\mathcal{H}_k$ using a single intensity measurement. To this end, we divide the SLM screen into distinct rows, encoding a different $\mathcal{H}_k$ in each row. 

In this configuration, the cylindrical lens performs a 1D Fourier transform independently and simultaneously on the Mattis problems encoded in each row.

To separate the optical intensity contributions of the different Mattis problems, the FPD technique superimposes a set of blazed gratings onto the SLM phase profile. This ensures parallel computation of several Hamiltonians with minimal crosstalk. Each Mattis energy is subsequently retrieved by measuring the optical intensity in spatially distinct regions of the camera plane.
More information can be found in Appendix~\ref{App-experimental}.

\section{Experimental results}
Our experimental demonstration uses the Wine classification dataset \cite{wine_109}, a standard task that categorizes samples into three distinct cultivars based on 13 chemical features. Its moderate dimensionality suits the current capabilities of our optical hardware. 

The inputs are normalized to the range $[-1, 1]$, and the targets are one-hot encoded in $\{-1, 1\}$. 
Continuous variables ($s_i$ and $\lambda_k$) are optimized by stochastic gradient descent, while binary patterns ($\xi_{ik}$) are updated using the Binary Optimizer (BOP) \cite{helwegen2019latent, laydevant2021training}.

The  number of SPIM evaluations required, per sample, per training step, is:
\begin{equation}
    n_\text{SPIM} = 2N_d \times (n_\text{free} + 2n_\text{nudge}) + 1 
\end{equation}
where $n_\text{free}=10$ is the number of steps taken during the inference phase, and $n_\text{nudge}=5$ is the number of steps taken during the nudged phase. 
To reduce runtime, we utilize focal plane division to compute two SPIM Hamiltonians 
in parallel. We limit the batch size to two to maintain a high signal-to-noise ratio given our macropixel size constraints. With these optimizations, training over 4 epochs (71 samples each) requires 45,440 total iterations, which takes approximately three hours.
 
\begin{figure}
    \centering
    \includegraphics[width=1\linewidth]{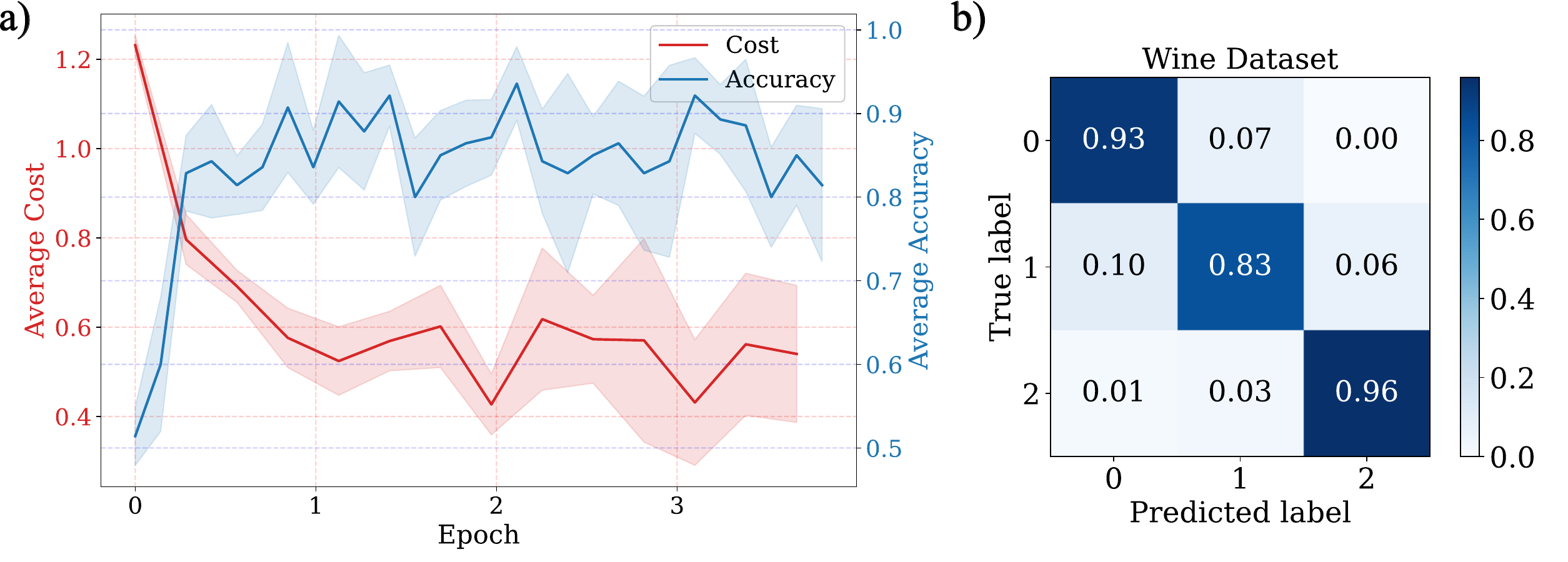}
    \caption{Experimental Results. (a) Average cost and accuracy over 7 independent training runs (4 epochs of 71 steps, 2 samples per step, moving average over 20 samples).  (b)  Average confusion matrix of 7 independently trained networks, evaluated on the test set.}
    \label{fig:exp_result}
\end{figure}

In Fig.~\ref{fig:exp_result}.a we report the average cost and accuracy over  seven independent training runs. Each training run consists of 4 epochs of 71 steps with 2 samples per step. To smooth out local fluctuations, we perform moving averages over 20 samples. We obtain an average test-set accuracy of $89.7 \pm 3.2\%$ (std, 7 runs), see Fig.~\ref{fig:exp_result}.b, to be compared with the simulation accuracy of $98.2 \pm 0.4\%$ (std, 10 runs). The discrepancy between the experiments and the simulations, as well as the fluctuations of the cost and accuracy after convergence, is a consequence of experimental noise. The system is susceptible to various imperfections, such as shot noise from the readout camera, phase fluctuations within the SLM, and laser power fluctuations. These perturbations directly affect SPIM energy estimation and can be viewed as an effective  temperature of the system \cite{pierangeli2020noise, pierangeli2020adiabatic}, which affects both gradient evaluation and the equilibrium state of the network.

After training, the energy landscape features distinct basins of attraction that guide the inference process to classify each sample.  
Interestingly, these basins are sufficiently deep and well-separated that inference can be accelerated by taking only a single gradient descent step with an optimized step size (rather than the $n_\text{free}=10$ steps used during training), yielding an average test-set accuracy of $85.71\pm 2.3\%$ (std, 7 runs). 
 
Additional numerical results using binary $\xi$ show:  1) that limiting to a rank $K = 3$ for $N_d = 8$ achieves optimal accuracy; and 2) that for this specific problem 4-bit neuron precision suffices for optimal accuracy and significantly speeds up inference.

\section{Numerical Results.}
Numerical studies show that directly scaling the above experimental architecture to larger datasets like MNIST is challenging because of the binary nature of the parameters $\bm \xi_{k}$, which limits expressivity, and because the Binary Optimizer \cite{helwegen2019latent} is difficult to use as it requires additional fine tuned hyperparameters.

However, future optical implementations should be able to encode continuous variables \cite{VEC_MAT-spall2020fully}, thereby eliminating these difficulties. Anticipating these developments, we  numerically evaluate our framework on the MNIST dataset (60k train, 10k test) when all parameters  are continuous. 
Interactions are all-to-all (contrary to standard implementations of EP, which generally use a layered architecture). The model is specified by the number of hidden neurons $N_d - 10$, which controls the strength of the nonlinearity in the dynamics, and the coupling rank $K$, which sets the number of independent parameters for a given $N_d$. We choose $N_d = 510$ and $K = 355$. The choice $K = 355$ corresponds to $\sim 70\%$ of the maximum rank $K=N_d=510$, which would allow fully independent all-to-all couplings.
 
We achieve a test-set accuracy of $97.81\pm 0.15\%$ (std, 10 runs). The continuous parameters and variables ($\bm{\xi}_k$, $\lambda_k$, and $s_i$) are  optimized via Adam \cite{kingma2014adam}, which is not necessary but accelerates convergence and reduces sensitivity to the learning rates (for inference and learning). Full hyperparameter details are provided in Appendix~\ref{App-C2}.
For comparison, a single hidden-layer network with $\sim 500 $ neurons (which has -to within $<0.12\%$- the same number of trainable parameters) trained using standard EP methods yields an accuracy of $\sim 97.5 \%$ as reported in \cite{scellier2017equilibrium}, $97.94 \pm 0.16 \%$ in \cite{ernoult2019updates} and $97.77\%$ on simulations of EP on the Kuramoto model \cite{rageau2025training}.

\begin{figure}
    \centering
    \includegraphics[width=0.98\linewidth]{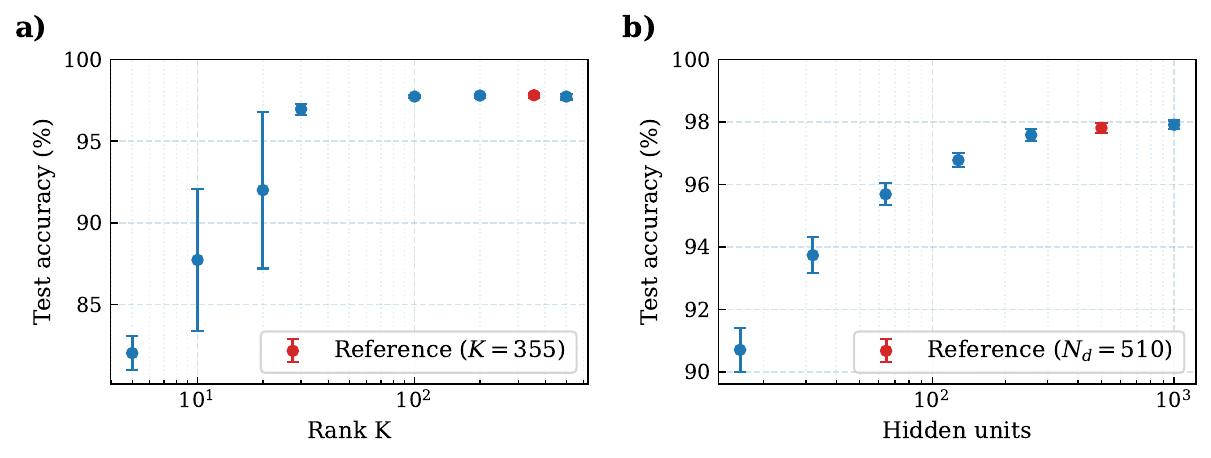}
    \caption{Numerical simulations of EP using SPIM on the MNIST dataset. (a) Test accuracy versus rank $K$ for $N_d = 510$. The maximum rank shown ($K = 700$) is larger than the theoretical maximum of $510$. (b) Test accuracy versus the number of hidden units ($N_d-10$), with rank scaling as $K \approx 0.7 N_d$. The horizontal axis is logarithmic in both plots.}
    \label{fig:rank_accuracy}
\end{figure}

To assess scalability for physical implementations, Fig.~\ref{fig:rank_accuracy} details the test accuracy scaling for the coupling rank $K$ [Fig.~\ref{fig:rank_accuracy}(a)] and the number of hidden units [Fig.~\ref{fig:rank_accuracy}(b)]. Notably, a minimal configuration ($N_d=32$, $K=22$) achieves $93.73 \pm 0.57\%$ accuracy, outperforming a linear classifier. Restricting unit values to 8-bit precision—consistent with the estimated resolution of a spatial light modulator (SLM)—accelerates inference without degrading accuracy (Table~\ref{tab:precision}). Note that the hyperparameters (\textit{e.g.}, hidden units, rank) are interdependent; see Appendix~\ref{App-E}, where further numerical results are reported. These results show that a physical implementation remains viable even with constrained resources.

\begin{table}[h]
    \caption{\label{tab:precision} Impact of units precision on accuracy and inference efficiency for MNIST ($N_d = 510$, $K = 355$). Inference Steps counts the number of steps needed for the prediction (the argmax of the outputs units) to stabilize.}
    \begin{ruledtabular}
    \begin{tabular}{ccc}
    Precision & Test Acc. (\%) & Inference Steps \\
    \colrule
    32-bit (float) & $97.68 \pm 0.18$ & $6.60 \pm 0.52$ \\
    12-bit & $97.84 \pm 0.17$ & $6.59 \pm 0.36$ \\
    10-bit & $97.50 \pm 0.24$ & $5.85 \pm 0.56$ \\
    8-bit & $97.67 \pm 0.15$ & $5.45 \pm 0.41$ \\
    6-bit & $96.61 \pm 0.24$ & $5.96 \pm 0.38$ \\
    4-bit & $90.60 \pm 1.16$ & $4.41 \pm 0.88$ \\
    \end{tabular}
    \end{ruledtabular}
\end{table}

Finally, as a proof of concept that the SPIM approach scales to more complex architectures, we train a network with one hidden layer in which the output units are not directly connected to the inputs, see Appendix \ref{App:Structured} for a detailed description of this structured architecture.
We obtain $97.71 \pm 0.10\%$ test accuracy on the MNIST dataset.

\section{Discussion.}
In this Letter, we proposed and demonstrated a hybrid optical-digital implementation of Equilibrium Propagation, exploiting a Spatial Photonic Ising Machine (SPIM) to replace numerically intensive computations with optical intensity measurements. This approach allows us to directly reuse the relatively simple, scalable, and well-developed platform of SPIMs—avoiding complex optical systems—to perform standard machine learning tasks. Moreover, because training relies directly on physical measurements (for both inference and $\lambda_k$ learning), we avoid the need for accurate, difficult-to-implement, digital twins. Our hybrid approach also offers practical flexibility compared to fully analog systems, such as the ability to use modern machine learning optimizers like Adam \cite{kingma2014adam}.

While SPIMs have previously been used to train Boltzmann machines on binary MNIST images \cite{yamashita2023low}, our implementation of EP proposes a distinct approach closer to modern machine learning. We use continuous variables, rely on deterministic evolution, and optimize a well-defined cost function. Our reinterpretation of the gauge transformation allows this to be implemented using binary patterns on existing SPIMs.

To fully exploit the potential of this method, future experimental implementations will need to scale up the SLM implementation, including performing the summation of the rank-1 Mattis problems optically (\textit{e.g.}, via wavelength multiplexing \cite{Li2023wavelength}), increasing the size of the SLM, reducing the size of the macropixels, and increasing the refresh rate. 

Ultimately, our results establish a concrete experimental foundation for a scalable optical-digital implementation of EP, potentially providing faster and energy-efficient training methods for physical machine learning.

\begin{acknowledgments}
DVA acknowledges the support of the French Community of Belgium through a FRIA fellowship. This research was supported by the FWO and F.R.S.-FNRS Excellence of Science (EOS) program grant 40007536. D.V., D.P, and C.C. acknowledge support from HORIZON EIC-2022-PATHFINDERCHALLENGES-01 HEISINGBERG Project no. 101114978.
\end{acknowledgments}

\newpage
\appendix

\section{Exact Finite Difference and Learning Rules}
\label{App-A}
The SPIM computes the energy:
\begin{equation}\label{eq:originalising}
    \mathcal{H} = - \frac{1}{2}\sum_{i,j} {J}_{ij} \sin(s_i) \sin(s_j),
\end{equation}
whose partial derivatives are given by:
\begin{equation}
    \label{eq:plain_derivative}
    \begin{split}
    \frac{\partial \mathcal{H}}{\partial s_m} &= - \cos(s_m) \sum_{i\ne m} {J}_{im} \sin(s_i) \\
    &\quad - {J}_{mm} \cos(s_m) \sin(s_m).
    \end{split}
\end{equation}

We evaluate these derivatives using a finite difference scheme. Shifting the phase $s_m$ by $\pm\pi/2$ or $\pm \pi/4$ while keeping the other phases fixed yields the expressions:
\begin{align}
    \mathcal{H}_{s_m + \pi/2} - \mathcal{H}_{s_m - \pi/2} 
    &= 2  \cos(s_m)\sum_{i\ne m} {J}_{im} \sin\left(s_i\right) , \label{Eq:Shiftpi/2}
   \\
        \mathcal{H}_{s_m + \pi/4} - \mathcal{H}_{s_m - \pi/4} &= -\sqrt{2}\sum_{i\ne m} \cos(s_m){J}_{im} \sin(s_i)  \nonumber \\
        & \ \quad - {J}_{mm} \cos(s_m) \sin(s_m).
        \label{Eq:Shiftpi/4}
\end{align}
The exact derivative can therefore be recovered via a linear combination of Eqs. (\ref{Eq:Shiftpi/2}) and (\ref{Eq:Shiftpi/4}). However, this requires four optical measurements. In order to halve the experimental run time, we sought to evaluate the partial derivative using only two optical measurements (\textit{i.e.}, one set of $\pm$ phase shifts). The simplest approach is to set the self-connections ${J}_{ii}$ to zero, which allows direct use of Eq. (\ref{Eq:Shiftpi/2}). However, as shown in Table~\ref{tab:accuracy_summary}, numerical simulations on the MNIST dataset demonstrate that omitting self-connections reduces stability and decreases final accuracy.

Consequently, we adopt a method that retains nonzero self-connections and uses a single $\pm \pi/4$ finite difference, but makes an approximation. We redefine the interaction energy, $\tilde{I}$, such that its gradient is given exactly by the $\pm \pi/4$ shift:
\begin{equation}
    \frac{\partial \tilde{I} }{\partial s_m}
    =
    \mathcal{H}_{s_m + \pi/4} - \mathcal{H}_{s_m - \pi/4} .
    \label{Eq:gradI}
\end{equation}
This corresponds to defining a new coupling matrix $\tilde{J}$ so that 
$\tilde I = - \frac{1}{2}\sum_{i,j} {\tilde J}_{ij} \sin(s_i) \sin(s_j)$. The original coupling matrix $J$
in Eq.~\eqref{eq:originalising} is related to $\tilde{J}$ by
\begin{equation}
    \label{eq:Jmodified_matrix}
    \tilde{J}_{im} = \left[ \frac{1}{\sqrt{2}} (1 - \delta_{im}) + \delta_{im} \right] J^{}_{im},
\end{equation}
which can be inverted as:
\begin{equation}\label{eq:Jdyn_infonction_J}
    J^{}_{im} = \left[ \sqrt{2} (1 - \delta_{im}) + \delta_{im} \right] \tilde{J}_{im}. 
\end{equation}
The dependence of the new interaction energy on the variational parameters $\bm{\xi}_k$ and $\lambda_k$ therefore takes the form
\begin{equation}
    \label{eq:modifiedE}
    \begin{split}
    \tilde{I} &= \frac{1}{2K} \sum_{k=1}^K \lambda_k \left[ (\sqrt{2} - 1) \sum_i \xi_{k,i}^2 \sin^2(s_i) \right. \\
    &\quad \left. - \sqrt{2} \left( \sum_i \xi_{k,i} \sin(s_i) \right)^2 \right].
    \end{split}
\end{equation}
The first term in the brackets isolates the modified self-coupling contribution, while the second is the standard Ising interaction.  (Note that for binary $\xi_{k,i} \in \{-1, 1\}$, 
the first term simplifies (since
$\xi_{k,i}^2 = 1$)
and can be reinterpreted as a modification of the self interaction which becomes $\frac{\alpha}{2}\lVert\bm{s}\rVert^2 + \frac{\alpha'}{2}\lVert{\sin (\bm s)}\rVert^2$ with $\alpha' = \frac{(\sqrt{2} - 1) }{K} \sum_{k=1}^K \lambda_k $).

Using the interaction energy  $\tilde{I}$ modifies the learning rules. Thus, the exact derivative with respect to $\xi_{k,m}$ becomes:   
\begin{equation}
    \begin{split}
        \frac{\partial \tilde{I}
        }{\partial \xi_{k,m}} &= \frac{\lambda_k}{K} \left[ (\sqrt{2} - 1) \xi_{k,m} \sin^2(s_m) \right. \\
        &\quad \left. - \sqrt{2} \sin(s_m) \sum_i \xi_{k,i} \sin(s_i) \right],
    \end{split}
\end{equation}
with a similar expression for $\partial \tilde{I}/\partial \lambda_k$.

In practice, rather than these exact expressions which cannot be evaluated optically, we use the  expressions Eqs. (\ref{eq:learnrule_lambda}, \ref{eq:learnrule_xi}) which are therefore approximate. There is thus a small incompatibility between the energy $E$ and the learning rules.

Numerical simulations presented in Table~\ref{tab:accuracy_summary} on the MNIST dataset show that using this approximate learning rule (denoted Approx LR in the table) does not compromise final accuracy.
Thus, we can halve the number of measurements, and therefore increase the speed of inference, without sacrificing accuracy.

\begin{table}[h]
    \caption{\label{tab:accuracy_summary} Numerical comparison of test accuracies on the MNIST dataset ($N_d = 500$, $K = 355$, continuous patterns $\bm \xi_k$) using different inference methods and learning rules. Three inference methods for $\frac{\partial E}{\partial s_m}$ are compared: (i) Original: uses the coupling matrix $J$ in Eq.~\eqref{eq:originalising}
    (which can be obtained using the appropriate linear combinations of $\pm\pi/2$ and $\pm\pi/4$ shifts  in Eqs.~\eqref{Eq:Shiftpi/2} and \eqref{Eq:Shiftpi/4}); (ii) No self-coupling: 
    uses the coupling matrix $J$ in Eq.~\eqref{eq:originalising} and
    sets $J_{ii}=0$ (which can be obtained using only $\pm\pi/2$ shifts Eq.~\eqref{Eq:Shiftpi/2}); 
    and (iii) Main: uses the effective coupling $\tilde J$ and energy $\tilde I$ (which can be obtained using 
    only $\pm\pi/4$ phase shifts Eq.~\eqref{Eq:Shiftpi/4}). For methods (ii) and (iii) 
    we compare the exact learning rule (Exact LR) which 
    computes the exact gradient with respect to the parameters $\lambda_k$ and $\xi_{k,i}$, and 
 the approximate learning rule (Approx LR) that uses
 Eqs. (\ref{eq:learnrule_lambda}, \ref{eq:learnrule_xi}).
 }
    \begin{ruledtabular}
    \begin{tabular}{lcc}
    Method & Test Acc. (\%) (std, 10) \\
    \colrule
    Original ($\pm \pi/2$ and $\pm \pi/4$)           & $97.47 \pm 0.22$  \\
    No self-coupling ($\pm \pi/2$, Approx LR)     & $48.77 \pm 20.04$ \\
    No self-coupling ($\pm \pi/2$, Exact LR)     & $96.68 \pm 0.33$ \\
    Main ($\pm \pi/4$, Approx LR) & $97.69 \pm 0.17$ \\
    Main ($\pm \pi/4$, Exact LR)  & $97.78 \pm 0.23$ \\
    \end{tabular}
    \end{ruledtabular}
\end{table}

\section{Structured Architectures}
\label{App:Structured}
In the implementation described in the main text, the coupling is symmetric and all-to-all by construction. While highly expressive, this full-rank matrix can induce a glassy energy landscape that traps the system in local minima and hinders learning. 
(This difficulty is well known in Hopfield networks that have a very similar Hamiltonian \cite{hopfield1982neural, amit1985storing}.)
The system described in the main text therefore lacks  the hierarchical advantages that deep networks typically hold over shallow ones \cite{mhaskar2016deep}.

To circumvent this, we note that the physical setup only requires all-to-all connectivity among neurons displayed \emph{simultaneously} within the same Mattis problem. By distributing the network, we can engineer deep, complex connectivities.
\begin{figure}[h]
    \centering
    \includegraphics[width=0.5\textwidth]{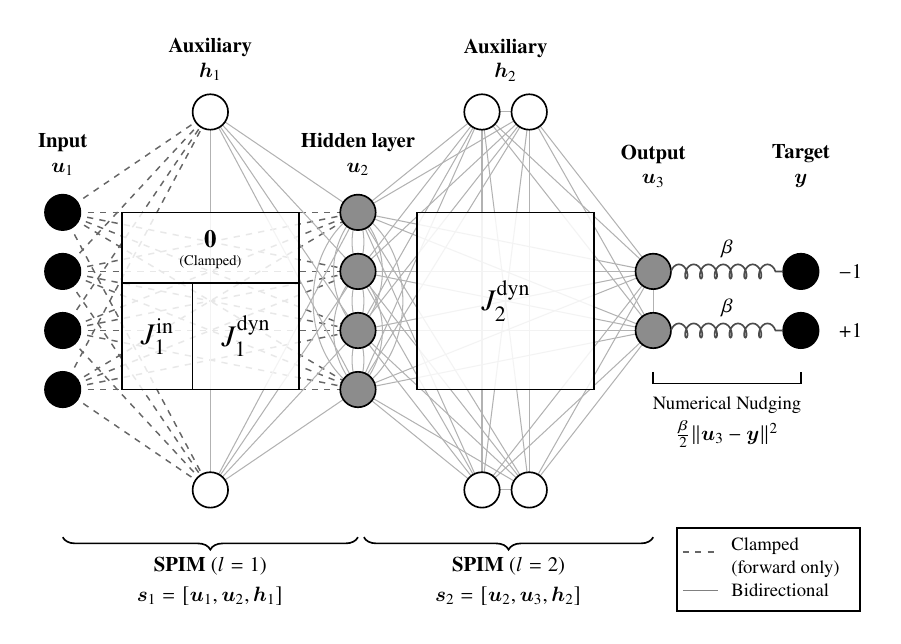}
    \caption{Two-layer structured architecture. The network uses two SPIMs ($l=1,2$). Clamped inputs $\bm{u}_1$ are projected via $J_1^\text{in}$, while remaining neurons couple through $J_{1,2}$. Outputs $\bm{u}_3$ are nudged toward targets $\bm{y}$.}
    \label{fig:fully_connected_with_layers}
\end{figure}

As a proof of concept, we introduce a structured layered architecture (Fig.~\ref{fig:fully_connected_with_layers}). Each intermediate layer $l$ features two sets of shared neurons, $\bm{u}_{l}$ and $\bm{u}_{l+1}$, connecting it to layers $l-1$ and $l+1$, respectively, alongside layer-specific auxiliary hidden neurons $\bm{h}_l$. Specifically, the states $\bm{u}_{l}$ represent the input neurons for $l=1$, the output neurons for $l=L+2$, and the layer of hidden neurons otherwise. At the level of a Mattis problem, this reads, for $l>1$:
\begin{equation}\label{eq:defphi}
    \bm{s}_l = \left[ \bm{u}_l, \bm{u}_{l+1}, \bm{h}_l \right] .
\end{equation} 
The energy functional therefore takes the form
\begin{equation}\label{eq:network_architecture}
\begin{split}
    E =\; & B(\bm{s}_1, \bm{u}_1) + \sum_{l=2}^{L+1} I_{l,l+1}(\bm{s}_l) + \frac{\beta}{2} \|\bm{u}_{L+2} - \bm{y}\|^2 \\
    & + \frac{\alpha}{2} \left( \sum_{l=2}^{L+2} \|\bm{u}_l\|^2 + \sum_{l=1}^{L+1} \|\bm{h}_l\|^2 \right).
\end{split} 
\end{equation}
which generalizes Eq.~\eqref{eq:EnergyFunction}. The theory for the single-layer setup can then be easily adapted to this case.

Preliminary numerical investigations show that because the shared hidden neurons $\bm{u}_{l}$ ($l \in \{2, \dots, L+1\}$) couple to two adjacent all-to-all systems, and given the initialization described in Appendix~\ref{App-C2}, they are sometimes prone to instability under the standard inference rules. However, we could stabilize them by adjusting their update rate during inference.

\section{Experimental Setup}
\label{App-experimental}
Here, we review the experimental setup and detail our adaptation of the gauge encoding method \cite{fang2021experimental}, which allows us to experimentally implement Equilibrium Propagation (EP) using binary $\xi_{k,i} \in \{-1,1\}$ patterns on the SLM setup. The  implementation used in the present work is identical to the fully programmable SPIM described in Ref.~\cite{veraldi2024fully}. 

SPIMs optically compute the energy of a set of spins $\bm{\sigma} \in \{-1, +1\}^N$ subject to an Ising Hamiltonian with a coupling matrix $J$. Using an SLM, the SPIM encodes the spins and couplings in the phase of a coherent optical field. This modulated field then propagates through a lens that performs a spatial Fourier transform. Interference at the focal plane gives a measured intensity proportional to the weighted sum of pairwise spin interactions, that is, the Ising energy \cite{pierangeli2019large}. Most SPIMs are limited to encoding rank-1 interaction $J = \bm{\zeta} \bm{\zeta}^\top$ (\textit{i.e.}, Mattis model) \cite{pierangeli2019large, fang2021experimental, yamashita2023low}. To extend the SPIM's capability, the Hamiltonian can be decomposed into a linear combination of Mattis-type problems using Eq.~\eqref{Eq:Mattis}, giving:
\begin{equation}
    H(\bm{\sigma}) = - \frac{1}{2}\sum_{k=1}^{K} \mathcal{H}_k = - \frac{1}{2} \sum_{k=1}^{K} \lambda_k\sum_{i,j=1}^{N}  \zeta_{k,i} \sigma_i \zeta_{k,j} \sigma_j,
\label{eq-app-spim}
\end{equation}
where $\lambda_k$ are numerical coefficients. The spins $\sigma_j = \exp(i \phi_j) = \pm 1$ are represented as local binary phase delays $\phi_j \in \{0, \pi\}$ and are encoded alongside the couplings $\zeta_{k,i}$ in the optical phase using the gauge transformation method \cite{fang2021experimental}. The intuition is to map binary spins to circular ones via a rotation around the $z$-axis by an angle $\alpha_{i,k} = \arccos(\zeta_{i,k})$ (where $\zeta_{i,k} \in [-1, 1]$). This gives an effective $z$-component $\sigma_{i,k}'^z = \zeta_{ik}\sigma_i$, allowing both the spin and the interaction coefficient to be encoded in a single variable (see Ref.~\cite{Li2023wavelength} for full details). Future implementations of EP could bypass these restrictions by employing a different SLM setup, which permits the optical realization of both spins and couplings as continuous variables \cite{VEC_MAT-spall2020fully}.

Our adaptation of this transformation identifies the spins $\sigma_i$ with our binary parameters $\xi_{k,i}$, and the angles $\alpha_{k,i}$ with the continuous units $s_i$. Consequently, each Mattis energy takes the form:
\begin{equation}
    \mathcal{H}_k \propto \sum_{i,j=1}^{N}\xi_{k,i} \xi_{k,j} \cos\left(s_i + \frac{\pi}{2}\right) \cos\left(s_j + \frac{\pi}{2}\right),
\end{equation}
where $s_i\in [-\pi/2, \pi/2)$.
This forms the basis of the framework introduced in Eqs.~\eqref{eq:quadraticInteraction} and \eqref{Eq:Mattis}. To ensure that nonlinearity vanishes at the origin, we encode a sine function by shifting $s_i$ by $\pi/2$. Alternatively, one could retain the cosine and shift the origin of the quadratic norms by $\pi/2$.

\begin{figure}
    \centering
    \includegraphics[width=1\linewidth]{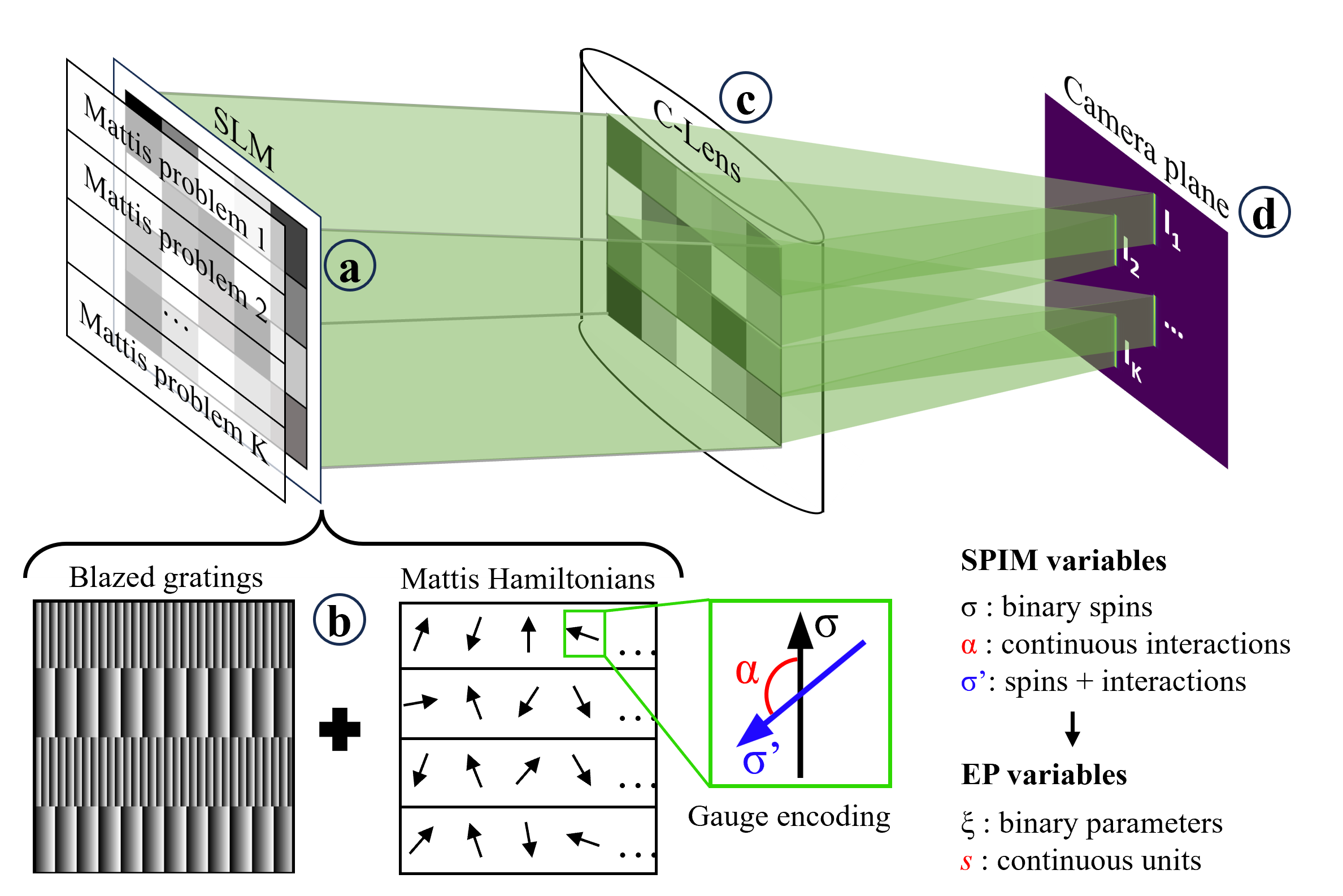}
    \caption{ Schematic of the focal plane division method to encode multiple Mattis problems simultaneously on a single SPIM. ({a}) The Spatial Light Modulator is used to encode a series of Mattis problems and blazed gratings onto the optical beam's phase profile, while ({b}) exploiting gauge encoding for spins and interaction coefficients. ({c}) A cylindrical lens focuses the propagating beam in the $x$-direction. The induced phase shift from the blazed gratings creates a set of distinct spots on the camera plane ({d}). The intensity of each spot provides the value of the corresponding Mattis Hamiltonian.}
    \label{fig:supp_exp}
\end{figure}

In practice, we employ the Focal Plane Division (FPD) technique \cite{veraldi2024fully} to simultaneously compute the energies of multiple $\mathcal{H}_k$ using a single intensity measurement. We divide the SLM screen into distinct rows, encoding a different $\mathcal{H}_k$ in each row, see Fig~\ref{fig:supp_exp}.a. The Mattis Hamiltonians are encoded into macropixels composed of $P_x \times P_y$ pixels; each macropixel carries information regarding both a spin and its associated coupling. 
Typical values are $P_x=30$ and $P_y=15$ which occupies an area of $0.074~\text{mm}^2$ on the SLM.
Specifically, the $i$-th macropixel of the $k$-th SLM row imparts a phase modulation defined by \cite{fang2021experimental}:
\begin{equation}
     \phi_{k,i}^l = \xi_{k,i} \frac{\pi}{2}+(-1)^ls_i,
    \label{eq:phase_modulation}
\end{equation}
where $l$ denotes the pixel index within a macropixel along the $x$-direction ($1 \le l \le P_x$). 
To separate the optical intensity contributions of the different Mattis problems, the FPD technique superimposes a set of blazed gratings onto the SLM phase profile, see Fig~\ref{fig:supp_exp}.b. This ensures parallel computation of several Hamiltonians with minimal crosstalk. The Ising energy is computed by propagating the light through a cylindrical lens aligned with the SLM's $x$-axis. In this configuration, the lens performs a 1D Fourier transform independently and simultaneously on the Mattis problems encoded in each row, see Fig~\ref{fig:supp_exp}.c.
Each Mattis energy is subsequently retrieved by measuring the optical intensity at spatially distinct regions of the camera plane, see Fig~\ref{fig:supp_exp}.d.

The experimental setup is schematized in Fig.~\ref{fig:SpimSetup}.b. A linearly polarized continuous-wave laser beam ($\lambda=532$~nm) is expanded to illuminate a phase-only SLM (Hamamatsu LCOS-SLM X15213S). The beam wavefront is modulated according to Eq. (\ref{eq:phase_modulation}) with 213 distinct phase levels within the $[0, 2\pi]$ interval. A cylindrical lens ($f_1=500$~mm, with the optical axis perpendicular to the optical table) focuses the modulated beam onto the sensor of a 12-bit CMOS camera (Basler a2A2590-60umPRO).

\section{All optical gradient evaluation }
\label{App-optical-grad}
Training the network via EP requires gradients of the energy functional with respect to the units $\textbf{s}$ and the parameters $\{ \lambda_k, \xi_{k,i} \}$, defined in Eqs.~\eqref{Eq:HSPIMpi/4},~\eqref{eq:learnrule_lambda},~\eqref{eq:learnrule_xi}, respectively. While optically computing the gradients for the units $\textbf{s}$ and parameters $\lambda_k$ is straightforward, the same does not hold true for the gradients with respect to the parameters $\xi_{k,i}$. Here we show that these gradients can in principle be computed optically.

From Eq.~\ref{eq-app-spim}, the energy computed by the SPIM (neglecting the $k$ index) is:

\begin{equation}
    \begin{split}
    \mathcal{H}_\text{SPIM} &\propto \sum_{ij} \xi_i \xi_j \rho(s_i)\rho(s_j) \\
    &= \left( \xi_i\rho(s_i) + \sum_{j\neq i} \xi_j \rho(s_j) \right)^2 \\ 
    &= \xi_i^2\rho(s_i)^2 + \left( \sum_{j\neq i} \xi_j \rho(s_j) \right)^2 \\
    &\quad + 2\xi_i \rho(s_i)\sum_{j\neq i} \xi_j \rho(s_j)
    \end{split}
\end{equation}

Consider two consecutive measurements of $\mathcal{H}_\text{SPIM}$, one with $\xi_i = 1$ and one with $\xi_i = -1$, and calculate their difference:

\begin{equation}
    \begin{split}
    \Delta\mathcal{H}_i &= \mathcal{H}_\text{SPIM}(\xi_i = 1) - \mathcal{H}_\text{SPIM}(\xi_i = -1) \\ 
    &= 4\rho(s_i)\sum_{j\neq i} \xi_j \rho(s_j)
    \end{split}
\end{equation}

Comparing this expression to the gradient of $\mathcal{H}$ with respect to $\xi_{k,i}$ in Eq.~\eqref{eq:learnrule_xi}, we have:

\begin{equation}
    \begin{split}
    \frac{\partial I}{\partial \xi_{k,i}} &\approx -\frac{\lambda_k}{K} \rho(s_i) \sum_j \xi_{k,j} \rho(s_j) \\
    &= \frac{\lambda_k}{4K} \left( \Delta\mathcal{H}_i^k + 4\sigma_i\xi_i\sigma_i \right)
    \end{split}
\end{equation}

Therefore, it is possible to calculate the gradient with respect to $\xi_{k,i}$ using two additional measurements and a single digitally computed term, which can reasonably be neglected in the large $N \gg 1$ limit. 

\section{Scaling to larger problems}
\label{app:scaling}
Below, we discuss scaling to larger datasets in future experiments. We consider asymptotic computational complexity, required SLM size, and whether input couplings should be evaluated numerically. 

 We recall that there are $N_i$ input units, $N_{out}$ output units, and $N_{hid}$ hiddent units. The total number of dynamical variables is $N_d = N_{hid} + N_{out} $. We generally have $N_i \gg N_d$.

\textbf{Computational complexity.} A single-step inference requires summing $2K$ light intensities (one per Mattis Hamiltonian) to evaluate the gradient $\frac{\partial \tilde{H}}{\partial s_m}$ for $N_d$ dynamical units. This yields an inference complexity of $\mathcal{O}(K N_d)$. For comparison, a naive digital evaluation of the low-rank gradient of the original interaction, Eq.~\eqref{eq:quadraticInteraction}, requires computing:
\begin{equation}
    \frac{\partial \mathcal{I}}{\partial s_m} = - \frac{2}{K} \cos(s_m) \sum_{k=1}^K \lambda_k \xi_{k,m} \left( \sum_{j=1}^{N_d} \xi_{k,j} \sin(s_j) \right)
\end{equation}
Precomputing the inner sum for all $K$ patterns costs $\mathcal{O}(N_d K)$, and subsequently evaluating the outer sum over the rank for all $N_d$ dynamical neurons similarly costs $\mathcal{O}(N_d K)$, giving the same single-step inference complexity of $\mathcal{O}(K N_d)$ (because the input projection $B(\bm{s}, \bm{u})$ can be precomputed, it does not affect this asymptotic scaling).

Therefore using the decomposition into Mattis problems yields the same asymptotic complexity as a digital implementations.
 However the optical implementation could be advantageous if 
the $K$ Mattis problems are summed optically, for instance by using wavelength multiplexing. Indeed, in this case 
the optical complexity reduces from $\mathcal{O}(K N_d)$ to
$\mathcal{O}(N_d)$.

During learning the optical implementation  presents an asymptotic advantage. Indeed, evaluating the gradient $\frac{\partial \mathcal{H}}{\partial \lambda_m}$ requires only $\mathcal{O}(K)$ measurements, while the digital evaluation of Eq.~\eqref{eq:learnrule_lambda} requires  $\mathcal{O}(K N_d^2)$ operations.

\textbf{SLM size and maximum rank.} Increasing the rank $K$ (the number of Mattis Hamiltonians) improves expressivity for a specific $N_d$ but demands more SLM pixels. Here, we show that maximum expressivity does not require a rank scaling with the total system size $N_d + N_i$ or the problem size $N_i$. Rather, the rank is bounded by the number of dynamical units $N_d$.

To see this, note that the coupling matrix given in Eq.~\eqref{Eq:SymmCoupling} can be modified to
\begin{equation}
    J = \begin{bmatrix}
    A  & J^\text{in} \\
    (J^\text{in})^\top & J^\text{dyn} 
    \end{bmatrix} ,
    \label{Eq:JA}
\end{equation}
where the input-input block $A \in \mathbb{R}^{N_i \times N_i}$ is arbitrary, since it
enters neither the dynamics nor the learning rule.
One easily shows that the maximum rank of a matrix of the form Eq.~\eqref{Eq:JA} with $A$ arbitrary is given by the dimension of $J^\text{dyn} $, that is by $N_d$.

This upper bound on the rank must be adjusted slighlty for binary implementations ($\xi_{k,i} \in \{-1, 1\}$ as used in the experiment). Because each coordinate acts as an independent sub-Gaussian Rademacher variable, subspace embedding theorems \cite{vershynin2018high} imply that a random binary dictionary of size $K_\text{binary} \gtrsim c N_d \log (N_i + N_d)$ with $c$ a constant approximates a continuous rank-$K$ matrix with high probability.
This suggests that the required rank in the binary case is only slightly larger than the maximum rank $N_d$ in the continuous case.

\textbf{Evaluating Input Couplings Digitally.}
Because $N_i \gg N_d$, most of the space on the SLM is taken up with encoding the inputs. 
At maximum rank, this requires $N_d(N_i + N_d)$ pixels.

To reduce the required SLM size, the optical evaluation can be restricted to the computationally intensive interaction $I(\bm{s})$ in Eq.~\eqref{eq:EnergyFunction}. The required SLM size is then reduced to $N_d^2$ pixels. In this configuration the  input projection $B(\bm{s}, \bm{u})$ would be evaluated digitally, modifying the gradient descent step of Eq.~\eqref{Eq:dEds} to:
\begin{equation}
    \begin{split}
        \frac{\partial E}{\partial s_m} &= - \left( (J^\text{in})^\top \rho(\bm{u}) \right)_m \rho'(s_m)  + \left.\frac{\partial I}{\partial s_m}\right|_{\text{SPIM}} \\
        &\quad + \alpha s_m + \beta (s_m - y_m) \delta_{m \in \text{out}}\ .
    \end{split}
    \label{Eq:dEds-B}
\end{equation}
The input projection bias $\left( (J^\text{in})^\top \rho(\bm{u}) \right)_m$ is precomputed once per input vector $\bm{u}$ and reused across all subsequent inference steps. As an added benefit, computing $B(\bm{s}, \bm{u})$ digitally provides a straightforward method to enforce specific structural constraints, such as setting direct input-output connections to zero.

\section{Initialization and Hyperparameters}
\label{App-C2}

The initialization of the parameters must be such that during the first steps of training, the system is stable and converges reasonably fast to an energy minimum. This only has to be imposed initially, as empirically we find that the system has a tendency to  increase stability during training.

To avoid instability during the initial steps, we initialize the parameters $\xi_{k,i}$ and $\lambda_k$ so that the matrix elements of $\tilde J$ have variance $\sigma^2_{\tilde J} = 1/N_d$. This ensures the average strength of the interneuron forces is of order 1 and independent of $N_d$. In addition, stability may be enhanced by increasing $\alpha$ in Eq.~\eqref{eq:EnergyFunction}.

We now compute $\sigma^2_{\tilde J}$. Using the effective weight matrix $\tilde{J} \in \mathbb{R}^{M \times M}$ defined in Eq.~\eqref{eq:Jmodified_matrix}, we have
\begin{equation}\nonumber
    \sigma^2_{\tilde J} = \text{Var}\left[\left( \frac{1}{\sqrt{2}} (1 - \delta_{ij}) + \delta_{ij} \right)\frac{1}{K}\sum_{k=1}^K \lambda_k \xi_{k,i} \xi_{k,j}\right].
\end{equation}
Assuming $\lambda_k$ and $\xi_{k,i}$ are independent with zero mean and respective variances $\sigma^2_\lambda$ and $\sigma^2_\xi$, we obtain for the off-diagonal terms:
\begin{equation}
    \label{eq:key_relation2} 
   \sigma^2_{\tilde J} = \frac{1}{2 K^2}\sum_k^{K} \text{Var}\left[\lambda_{k}\right] \text{Var}\left[\xi_{k,i} \xi_{k,j}\right] = \frac{\sigma^2_\lambda (\sigma^2_\xi)^2}{2K}   .
\end{equation}
The diagonal elements have a variance that is twice as large. We impose $ \sigma^2_{\tilde J} = 1/N_d$ for the off-diagonal elements only. (Indeed, for large matrix sizes, Wigner's Semicircle Law dictates that the largest eigenvalues, and thus the stability of the dynamics, do not depend on the diagonal variance \cite{bai1988necessary, tao2023topics}).

The relation in Eq.~\eqref{eq:key_relation2} offers freedom in choosing $\sigma_\lambda$ and $\sigma_\xi$. Because the patterns $\bm{\xi}_k$ represent physical states, they must be strictly bounded. In the binary case, they are Rademacher variables ($\pm 1$), yielding $\sigma^2_\xi = 1$. Therefore, in this case we take:
\begin{equation}
    \lambda_{k}  \sim \mathcal{N}\left(0, 2\frac{K}{N_d}\right).
\end{equation}

In the  continuous case,  the ${\xi}_{k,i}$ 
are physically constrained within $(-1, 1)$. Choosing a uniform distribution $\xi_{k,i} \sim \mathcal{U}(-0.9, 0.9)$ yields $\sigma^2_\xi = 0.27$. Whereupon, we take
\begin{equation}
    \lambda_{k}  \sim \mathcal{N}\left(0,\frac{K}{0.03645 N_d}\right).
\end{equation}

The rest of the network parameters and hyperparameters are initialized empirically, see Table~\ref{tab:Exp Hyper2}.

\begin{table*}[t]
    \label{tab:Exp Hyper2}
    \caption{Experimental and Numerical Hyperparameters.}
    \begin{ruledtabular}
    \begin{tabular}{lccc}
     & \textbf{Exp. (all-to-all)}  & \textbf{Num. (all-to-all)}& \textbf{Num. (layered all-to-all)} \\
     & \textit{Wine, binary $\xi$} & \textit{MNIST, continuous $\xi$} & \textit{MNIST, continuous $\xi$}\\ 
    \colrule
    Input ($N_i$) & 13 & 784 & 784 \\
    Hidden ($N_h$) & 5 & 500 \textit{(default)} & 400-100 \\
    Hidden layer & \textit{n.a.} & \textit{n.a.} & 500 \\
    Output ($N_o$) & 3 & 10 & 10 \\
    Rank ($K$) & 20 & 355 \textit{(default)} & 800-400\\
    Nudging ($\beta$) & 0.9 & 0.75 & 0.9 \\
    Potential ($\alpha$) & 2 & 2 & 2 \\
    Epochs ($n_\text{epoch}$) & 4 & 50 & 150\\
    Free steps ($n_\text{free}$) & 10 & 40 & 48\\
    Nudged steps ($n_\text{nudge}$) & 5 & 10 & 12\\
    Batch size & 2 & 64 & 64\\
    Train/Test split & 80\% / 20\% & 60000 / 10000& 60000 / 10000 \\
    Inference learning rate & 0.05 & 0.12 & 0.12 \\
    Learning rate ($\lambda_k$ / $\xi_{k, i}$) & 0.02 / \textit{n.a.} & 0.10 & 0.08 \\
    Optimizer ($\lambda_k$ / $\xi_{k, i}$) & SGD / BOP & ADAM & ADAM \\
    BOP ($\tau$ / $\gamma$) & $5 \times 10^{-8}$ / $1 \times 10^{-4}$ & \textit{n.a.} & \textit{n.a.} \\
    ADAM ($\beta_1$ / $\beta_2$ / $\hat{\epsilon}$) & \textit{n.a.} & 0.9 / 0.999 / $10^{-8}$ & 0.9 / 0.999 / $10^{-8}$ \\
    L2 regularization ($\lambda_k$ / $\xi_{k, i}$) & 0.001 / \textit{n.a.} & 0.00 & 0.00 \\
    \end{tabular}
    \end{ruledtabular}
\end{table*}

\section{Supplementary Numerical Results}
\label{App-E}
We present additional numerical results on the MNIST dataset with continuous patterns. 

\textbf{Trade-off between hidden units and rank.}
Here, we discuss the relative impact of the rank versus the number of hidden units on the accuracy of the network. 
To isolate the impact of these two parameters, we compare networks with different $N_h$ and $K$ but the same  total number of independent parameters $P$.  
The results are presented in Table~\ref{tab:const_P}.

For the range of parameters shown, increasing the number of parameters $P$ always increases performance. For constant $P$, decreasing the rank $K$ and therefore increasing the number of hidden variables improves performance.
Thus there is a trade-off between the number of hidden units $N_h$ (which increases network non-linearity) and the rank $K$ (which increases the expressivity of the coupling).

The relation between the number of hidden units $N_h$ and the total number of independent parameters $P$ is obtained as follows. The dimension of the manifold $\mathcal{S}_{N_i + N_d,K}$ of rank-$K$ symmetric matrices of size $N_i + N_d$ given by \cite{daniilidis2016spectral}:
\begin{equation}
  \text{dim}(\mathcal{S}_{N_i + N_d, K}) = K(N_i + N_d-K) + \frac{K(K+1)}{2}.
\end{equation}
Setting $P = \text{dim}(\mathcal{S}_{N_i + N_d, K})$, we can solve
 for $N_d$ and $K$, to obtain the required number of hidden units for given $P,K, N_i, N_o$:
\begin{equation}
    N_h = \frac{P}{K} + \frac{K-1}{2} - N_i - N_o .
\end{equation}

\begin{table*}[t]
\caption{\label{tab:const_P} Numerical simulations of performance on MNIST dataset for different network configurations. $P$ is the total
 number of independent parameters, $K$ the rank of connection matrix, and $N_d$ the number of hidden plus output units.
 }
\begin{ruledtabular}
\begin{tabular}{lcccccc}
$P$ & Rank $K$ & $N_d$ (hidden + output) & Train Acc. (\%) & Test Acc. (\%) \\
\colrule
20,700 & 23 & 127 & $96.41 \pm 0.82$ & $95.28 \pm 0.73$ \\
       & 24 & 90  & $95.93 \pm 0.52$ & $94.89 \pm 0.49$ \\
       & 25 & 56  & $95.27 \pm 0.30$ & $94.41 \pm 0.32$ \\
\colrule
33,345 & 30 & 342 & $98.27 \pm 0.39$ & $96.79 \pm 0.27$ \\
       & 38 & 112 & $97.06 \pm 0.42$ & $95.92 \pm 0.45$ \\
       & 39 & 90  & $96.58 \pm 0.29$ & $95.51 \pm 0.28$ \\
\colrule
39,997 & 37 & 315 & $98.73 \pm 0.27$ & $97.09 \pm 0.23$ \\
       & 46 & 108  & $97.23 \pm 0.31$ & $95.99 \pm 0.26$ \\
       & 47 & 90  & $96.97 \pm 0.29$ & $95.77 \pm 0.22$ \\
\end{tabular}
\end{ruledtabular}
\end{table*}

\bibliography{apssamp}

\end{document}